\documentstyle[aps,multicol,pre,epsf]{revtex}
\begin{document}
\title{The Effect of Shear on Phase-Ordering Dynamics with 
Order-Parameter-Dependent Mobility: The Large-n Limit}
\author{N. P. Rapapa}
\address{Department of Physics and Astronomy, The University,
Manchester M13 9PL, UK}
\date{\today}
\maketitle

\begin{abstract}
The effect of shear on the ordering-kinetics of a conserved order-parameter 
system with $O(n)$ symmetry and order-parameter-dependent mobility 
$\Gamma({\,\vec\phi}\,) \propto (1- {\vec\phi}\,^2/n)^\alpha$ is studied 
analytically within the large-$n$ limit. In the late stage, the structure 
factor becomes anisotropic and exhibits multiscaling behavior with 
characteristic length scales $(t^{2\alpha+5}/\ln t)^{1/2(\alpha+2)}$ 
in the flow direction and $(t/\ln t)^{1/2(\alpha+2)}$ in directions 
perpendicular to the flow. As in the $\alpha = 0$ case, the structure 
factor in the shear-flow plane has two parallel ridges. 
\end{abstract}



\begin{multicols}{2}
\section{Introduction}
The dynamics of the ordered phases when a system is quenched from the  
high temperature phase into the low temperature region of two- or 
more-ordered phases has been of intense interest\cite{Bray}. It is now 
well established that in the late stage, both  the equal time pair 
correlation function $C({\bf r},t)$ and 
the structure factor $S({\bf k},t)$ obtained by the Fourier transform of 
$C({\bf r},t)$ obeys standard scaling. By standard scaling it is meant that 
$C({\bf r},t)$ and  $S({\bf k},t)$ can be written as $f(r/L)$ and 
$L^d g(kL)$ respectively, where $f(r/L)$ and $g(kL)$ are the scaling 
functions and $L(t)$ is the characteristic length scale in the system. For 
scaling to hold, $L(t)$ must be well separated from other length scales that 
may be present in the system. In most of the systems undergoing 
phase-ordering kinetics, the characteristic length scale $L(t)$ has a 
power-law dependence on the time $t$ elapsed since the quench, 
$L(t)\sim t^{1/z}$. The growth exponent $z$ depends on whether or not the  
order parameter is conserved. In the absence of shear, but with a constant 
mobility $\Gamma$, for nonconserved order parameter systems $z = 2$ and for 
conserved order parameter with no hydrodynamic  effects $z = 3$ (for 
scalar fields) and 4 (for vector fields). The characteristic length scale 
$L(t)$ is normally associated with the wavevector $k_m$ at which the 
spherically symmetric structure factor $S({\bf k},t)$ is maximum 
(i.e. $k_m \sim L^{-1}$).

When the phase-separating system is subjected to uniform shear\cite{Onuki}, 
the isotropy in the structure factor is broken as the domains grow faster 
in the flow direction than in directions perpendicular to the flow. 
This anisotropy is confirmed by 
simulation\cite{Rothman,Padilla,Yeomans,Gonnella} 
analytical\cite{Rapapa}, numerical\cite{Corberi} and 
experimental results\cite{Beysens,Hashimoto,Lauger}. 
We have previously shown analytically that within the large-$n$  
limit\cite{Rapapa}, the structure factor exhibits multiscaling with 
characteristic lengths $k_{my}^{-1} = k_{mz}^{-1}\sim (t/\ln t)^{1/4}$ 
and $k_{mx}^{-1} \sim (t^5/\ln t)^{1/4}$ in directions perpendicular to the 
flow and parallel to the flow respectively. For the scalar case (without 
hydrodynamic effects), renormalization group arguments\cite{Gonnella} 
predicts $k_{my}^{-1}\sim t^{1/3}$ and $k_{mx}^{-1} \sim t^{4/3}$, for 
the 'viscous hydrodynamic' regime L\"auger {\it et al.}\cite{Lauger} 
found experimentally $k_{my}^{-1} \sim t$ and $k_{mx}^{-1} \sim t^2$. 
The ratio $k_{mx}/k_{my} \sim 1/t$, is consistent with analytical, 
numerical, simulational and experimental results. 

In the case of unsheared phase-separation, there has been much study of 
the effects of an order-parameter-dependent 
mobility\cite{Langer,Jasnow,Puri,Castellano,Emmbra,Emmott,Ahluwalia}. It 
was suggested that a mobility of the form $\Gamma(\phi) =  1- \phi^2$ 
is appropriate for deep quenches\cite{Langer} and to account for the effects 
of external field such as gravity\cite{Jasnow}. Simulational 
calculations\cite{Puri} performed on phase-separation with scalar order 
parameter and mobility $\Gamma(\phi) =  1- a\phi^2$, show that for $a = 1$, 
$z = 4$ instead of 3 which is a result for constant mobility. A 
crossover from $z = 4$ to 3 was found for the case where  $a < 1$ 
(with $a > 0$)\cite{Puri}. For an $n$-vector order parameter, the simulation 
results\cite{Castellano} done with $\Gamma(\,{\vec\phi}\,) =  
1- a{\vec\phi}\,^2/n$ for $n = 2$, 3 and 4, shows a crossover from $z = 6$ 
to $z = 4$ for  $a < 1$, while for $a = 1$, $z = 6$. Emmott {\it et al.}
\cite{Emmbra,Emmott} considered a more general expression for the mobility 
given by 
\begin{equation}
\Gamma(\,{\vec\phi}\,) = \Gamma_0 \,(1- {\vec\phi}\,^2/n)^\alpha\,,
\label{mobe}
\end{equation}
where $\alpha \in \Re^+$. For scalar fields\cite{Emmbra} in the 
Lifshitz-Slyozov limit, they found $z = 3 + \alpha$ while for conserved 
vector fields within the large-$n$ limit\cite{Emmott}, multiscaling was found 
with two length scales $t^{1/2(\alpha+2)}$ and  ${k_m}^{-1} 
\sim (t/\ln t)^{1/2(\alpha+2)}$.  A different form for the  mobility, 
$\Gamma(\phi) = 1/(1 + \exp(\,\alpha\phi - \beta\phi^2)\,)$, where both 
$\alpha$ and $\beta$ are positive (with $\beta > \alpha$) was used by 
Ahluwalia\cite{Ahluwalia} in simulating the Cahn-Hilliard model of 
phase-separation. The domain patterns were found to be similar to the ones 
observed in viscoelastic phase separation. It is clear that different forms 
of mobility can be used depending on the type of problem concerned.

In this paper we are mainly concerned with the dynamics of the large-$n$ 
conserved order parameter with the simplest shear flow following a quench 
from the high temperature phase to zero temperature, and the case with 
an order-parameter-dependent mobility given by (\ref{mobe}). We show that 
in the late stage, the structure factor  becomes anisotropic and shows 
multiscaling behavior\cite{Coniglio} 
(i.e. $S({\bf k},t) \sim (\,(L_xL_yL_z)^{\varphi(k_xL_x, k_yL_y,k_zL_z)}\,)$ 
with characteristic length scales 
$L_x \sim k_{mx}^{-1} \sim (t^{2\alpha+5}/\ln t)^{1/2(\alpha+2)}$, 
$L_y \sim k_{my}^{-1} \sim (t/\ln t)^{1/2(\alpha+2)}$, and $L_z 
\sim k_{mz}^{-1} \sim (t/\ln t)^{1/2(\alpha+2)}$ extracted from 
the maxima of the structure factor. In the $k_z = 0$ plane, there are 
two parallel ridges, whose height and length depends on $\alpha$. 
These parallel ridges have been observed in experiments\cite{Beysens} for 
scalar fields in the absence of shear with $\alpha = 0$. It is 
worth mentioning that multiscaling is believed to be an artifact of the 
large-$n$ approximation as in the case of constant mobility and 
zero shear\cite{Humayun}. For systems with finite $n$, we expect scaling to 
be recovered asymptotically, although multiscaling may be exhibited as a 
preasymptotic effect\cite{Ruggiero}.

The paper is organised as follows: In the next section, model equations which 
take into the account both shear and non constant mobility are introduced. 
In section III, an exact solution for the structure factor is obtained in the 
scaling limit. The discussion of the results for specific values of 
$\alpha$ (i.e $\alpha = 1$ and 2) are presented in section IV. Concluding 
remarks are given in section V. 

\section{Model Equations}
In order to study the phase-separating system, the Cahn-Hilliard equation 
(generalised to $n$-vector order parameter ${\vec\phi}$) is given by
\begin{equation}
\partial_t{\vec\phi} = -{\bf\nabla}\cdot\left[\,\Gamma({\vec\phi}){\bf\nabla}
\left(-{\bf\nabla}^2{\vec\phi} + {\vec\phi} - ({\vec\phi}\,^2/n)\,{\vec\phi}
\,\right)\right].
\label{CAHE}
\end{equation}       
We are interested in a system with uniform shear flow which has a velocity 
field of the form ${\bf v} = \gamma y \,{\bf e_x}$, where $\gamma$ is the 
constant shear rate and ${\bf e_x}$ is a unit vector in the flow direction. 
For an incompressible system in the presence of shear, the term 
$({\bf v}\cdot{\bf\nabla}){\vec\phi}$ is added on the left hand side of 
(\ref{CAHE}) leading to 
\begin{equation}
\partial_t{\vec\phi}  + \gamma y \partial_x{\vec\phi} = -{\bf\nabla}\cdot\left
[\Gamma({\vec\phi}){\bf\nabla}\left(-{\bf\nabla}\,^2{\vec\phi} + {\vec\phi} - 
({\vec\phi}\,^2/n){\vec\phi}\,\right)\right].
\label{CAHES}
\end{equation}  

In the limit $n \rightarrow \infty$, ${\vec\phi}^2/n$ is replaced by its 
average in the usual way, and eq.(\ref{CAHES}) reduces to a linear 
self-consistent equation whose Fourier transform is given by 
\begin{equation}
\frac{\partial\phi_{\bf k}}{\partial t} 
- \gamma k_x\frac{\partial \phi_{\bf k}}{\partial k_y} 
= -{\bf k^2}a(t)^{\alpha}\,[{\bf k^2} - a(t)]\,\phi_{\bf k},
\label{CAHEF}
\end{equation}
where $\Gamma_0$ has been absorbed into the time scale, $\phi$ is (any) 
one component of $\vec\phi$, and $a(t) = 1 - \langle \phi^2\rangle$. 
Eq.(\ref{CAHEF}) has recently been solved numerically within a 
'self-consistent one-loop' approximation for a scalar order parameter in 
two dimensions by Gonnella {\it et al.}\cite{Suppa}. We believe that the 
oscillations (whose amplitudes decreases quite rapidly as $\alpha$ 
increases) found in\cite{Suppa} are slowly-decaying preasymptotic 
transients.
\section{Exact Solution in the Scaling Limit}
Eq.(\ref{CAHEF}) is a first order linear partial differential equation 
which can be easily solved by change of variables: 
$(k_x,k_y,t) \rightarrow (k_x,\sigma,\tau)$, with $t = \tau$ and 
$\sigma = k_y + \alpha k_x \tau$. With this transformation the left 
hand side of (\ref{CAHEF}) becomes $\partial\phi_k/\partial\tau$, and 
straight forward integration gives (after transforming back to original 
variables) $\phi_{\bf k}(t) = \phi_{\bf k}(0)\exp f({\bf k},t)$, where
\begin{eqnarray}
f({\bf k},t) &=& -(\kappa^2 + \mu^2)^2\,b_0(t) + 4\gamma k_x\mu 
\,(\kappa^2 + \mu^2)\,b_1(t)\nonumber\\
& & - 2\gamma^2 k_x^2 \,(\kappa^2 + 3\mu^2)\,b_2(t)+ 4\gamma^3 k_x^3\,\mu 
\,b_3(t) \nonumber\\
& & - \gamma^4 k_x^4 \,b_4(t) + (\kappa^2 + \mu^2)\,p_0(t) - 2 \gamma k_x 
\mu \,p_1(t)\nonumber\\ 
&  & + \gamma^2 k_x^2 \,p_2(t),  
\label{CAHS}
\end{eqnarray}
with $\kappa^2 = k_x^2 + k_z^2$, $\mu = k_y + \gamma k_x t$, 
$b_m(t) = \int_0^t dt'\,t'^m \,a(t')^\alpha$, 
$p_m(t) = \int_0^t dt'\,t'^m \,a(t')^{\alpha + 1} $.

From dimensional analysis, it is easy to see that to leading order in $t$: 
$a(t)\sim t^{-1/(\alpha + 2)}$, $L_y  \sim L_z \sim t^{1/2(\alpha + 2)}$ and 
$L_x \sim t^{(2\alpha + 5)/2(\alpha + 2)}$, with the dominant part of the  
$k_x$-dependence coming from the shear terms. In fact there are logarithmic 
corrections to these power law relations as we will see. It is reasonable 
to make the ansatz $a(t)\sim (\ln t/t)^{1/(\alpha + 2)}$ (\,this is true 
for $\gamma = 0$\cite{Emmott}\,) in the large-$t$ limit. Then to leading 
order in $t$, we have 
\begin{eqnarray}
b_m(t) &=& \frac{2\,t^m \,b_0(t)}{m \alpha + 2 m + 2}\,,\nonumber\\ 
p_m(t) &=& \frac{t^m \,p_0(t)}{m \alpha + 2m + 1}\,.  
\label{bopor}
\end{eqnarray}
Substituting (\ref{bopor}) in eq. (\ref{CAHS}) after making the following 
change of variables, 
\begin{equation}
\gamma k_x = \sqrt{\frac{p_0(t)}{t^2 b_0(t)}}\,u\,, 
\,\,\,k_y = \sqrt{\frac{p_0(t)}{b_0(t)}}\,v\,, 
\,\,\,k_z = \sqrt{\frac{p_0(t)}{b_0(t)}}\,w\,,
\label{chals}
\end{equation}
the structure factor $S({\bf k},t) = \langle \phi_{\bf k}(t)
\phi_{-\bf k}(t) \rangle$ becomes
\begin{eqnarray}
S({\bf k},t) &=& \Delta\exp\left(2\,\frac{p_0^2}{b_0}F(u,v,w)\right)\,,
\nonumber\\
F(u,v,w) &=& -\frac{u^4}{5}A_0(\alpha) - u^3 v A_1(\alpha) - 
2 u^2 v^2 A_2(\alpha)\nonumber\\ 
& & - 2 u v^3 A_3(\alpha) - v^4 + 
\frac{8}{15} u^2 A_5(\alpha) \nonumber\\
& & + \frac{4}{3} u v A_4(\alpha) + v^2 + w^2 - w^4 - 2 v^2 w^2 \nonumber\\
& & - \frac{2}{3}u^2 w^2 A_2(\alpha)- 2 u v w^2 A_3(\alpha)\,,
\label{FUS} 
\end{eqnarray}
where contributions to $F$ which vanish as $t \rightarrow \infty$ (at fixed 
$u,v,w$) have been dropped, $\Delta$ is the size of the initial fluctuation,  
$\langle \phi_i({\bf r},0)\phi_j({\bf r'},0) \rangle = 
\Delta\delta_{ij}\delta({\bf r - r'})$, and 
\begin{eqnarray}
A_0 &=& 5 -\frac{40}{\alpha + 4}+ \frac{60}{2\alpha + 6}
-\frac{40}{3\alpha + 8} + \frac{10}{4\alpha + 10}\,,\nonumber\\
A_1 &=& 4 -\frac{24}{\alpha + 4}+ \frac{24}{2\alpha + 6}
-\frac{8}{3\alpha + 8}\,, \nonumber\\
A_2 &=& 3 -\frac{12}{\alpha + 4}+ \frac{6}{2\alpha + 6}\,,\nonumber\\
A_3 &=& 2 - \frac{4}{\alpha + 4}\,, \nonumber\\
A_4 &=&\frac{3}{2} - \frac{3}{2\alpha + 6}\,,\nonumber\\
A_5 &=& \frac{15}{8} -\frac{30}{8\alpha + 24}+ \frac{15}{16\alpha + 40}\,,  
\label{AA}
\end{eqnarray} 
with $A_m(\alpha = 0) = 1$. In order to find the exact form of $b_0$ and 
$p_0$ in the large-$t$ limit, we consider the self-consistent equation 
for $a(t)$ given by 
\begin{eqnarray}
a(t) & = & 1 - \int \frac{d^3k}{(2\pi)^3}\,S({\bf k},t) \nonumber\\
& = &1 - \frac{\Delta}{(2\pi)^3\gamma t}\left(\frac{p_0}{b_0} \right)^{3/2} 
\nonumber\\
& & \times \int du\,dv\,dw\, \exp\left(2\,\frac{p_0^2}{b_0}\,F(u,v,w)\right). 
\label{self-cons}
\end{eqnarray}    
The above integral can easily be evaluated by the method of steepest 
descent using the points of global maxima in $F(u,v,w)$ provided 
$p_0^2/b_0 \rightarrow \infty$ as $t \rightarrow \infty$ ( this is the 
case we assumed before), therefore eq.(\ref{self-cons}) becomes
\begin{equation}
1 = \frac{C_1(\alpha)\Delta}{\gamma t p_0^{3/2}}\exp
\left(2\,\frac{p_0^2}{b_0}\,F_m(\alpha)\right)
\label{SECOS}
\end{equation}
where $C_1(\alpha)$ is a constant, $F_m(\alpha)$ is the value at the 
maxima $(\,u_m(\alpha),v_m(\alpha),w_m(\alpha)\,)$. The assumption 
that $a(t) \ll 1$ for $t \gg  1$ has also been used. 
\end{multicols}
\medskip

\begin{center}
\begin{tabular}{|c|c|c|c|c|c|}
\hline
\multicolumn{6}{|c|}{$\alpha = 1$}\\
\hline
Position (u,v,w) & Number  & F & Value & Type (3D) & Type (2D) \\ 
\hline
$(0,0,0)$ & 1 &   0         & 0 & Min & Min \\ \hline
$\pm(\sqrt{55/6}/3,0,0)$ & 2 & 55/168 & .32738 & IS & IMax \\ \hline
$\pm((33 - \sqrt{165})/18\sqrt{2},-1/\sqrt{2},0)$ &2& 
$31/168 - \sqrt{55/3}/28$ & .03160 & S2 & S \\ \hline
$\pm((33 + \sqrt{165})/18\sqrt{2},-1/\sqrt{2},0)$ &2& 
$31/168 + \sqrt{55/3}/28$ &.33744 & S1 & Max \\ \hline 
$\pm(0,0,1/\sqrt{2})$ & 2& 1/4 & .25 & S1 & - \\ \hline
$\pm(\sqrt{22}/3,-\sqrt{11/2}/4,\pm\sqrt{5/2}/4)$ &4& 39/112 & .34821 
&Max & - \\ 
\hline
\multicolumn{6}{|c|}{$\alpha = 2$}\\
\hline
Position (u,v,w) & Number  & F & Value & Type ($3D$) & Type $(2D)$ \\ 
\hline
$(0,0,0)$ & 1 &   0 & 0 & Min & Min \\ \hline
$\pm(\sqrt{7/8},0,0)$ & 2 & 14/45 & .31111 & IS & IMax \\ \hline
$\pm((7-\sqrt{7})/4\sqrt{2},-1/\sqrt{2},0)$ &2& 
$31/180 - \sqrt{7}/18$ & .02524 & S2 & S \\ \hline
$\pm((7 + \sqrt{7})/4\sqrt{2},-1/\sqrt{2},0)$ &2& 
$31/180 + \sqrt{7}/18$ & .31921 & S1 & Max \\ \hline 
$\pm(0,0,1/\sqrt{2})$ & 2& 1/4 & .25 & S1 & - \\ \hline
$\pm(\sqrt{35}/4,-\sqrt{7/5}/2,\pm \sqrt{3/5}/2)$ &4& 59/180 & .32778 
&Max & - \\ 
\hline
\end{tabular}
\end{center}

\underline{Table 1.} Stationary points of $F(u,v,w)$: Max = maximum, 
Min = minimum, S = saddle point (2D), S$n$ = saddle point of type $n$ (the 
matrix of second derivatives has $n$ positive eigenvalues), IS = `inflection 
saddle point' (one positive, one zero, one negative eigenvalue), 
IMax = `inflection maximum' (one zero, one negative eigenvalue). See also 
FIG. 1. for $2D$ results.

\medskip

\begin{multicols}{2}
Therefore to leading logarithmic accuracy, (\ref{SECOS}) leads to
\begin{eqnarray}
p_0^2(t) &\approx& \frac{b_0\ln t}{2 F_m} +  
\frac{3 b_0\ln b_0}{8 F_m}\,,\nonumber\\
 & & \nonumber\\
2\,p_0\,\frac{dp_0}{dt} & \approx & \frac{db_0}{dt}\left[ \frac{\ln t}
{2 F_m} + \frac{3 \ln b_0}{8 F_m}\right].
\label{syste}  
\end{eqnarray}
Using (\ref{syste}) and the relations $db_0/dt = a^ \alpha$, 
$dp_0/dt = a^ {\alpha + 1}$ we get 
\begin{equation}
a(t) \approx  \left(\frac{\ln b_0}{b_0}\right)^{1/2}\left[\frac{\ln t}
{8 F_m \ln b_0} + \frac{3}{32 F_m}\right]^{1/2}, 
\label{ATEQ}
\end{equation}
from which $b_0$ (since $db_0/dt = a^\alpha$) is found to be 
\begin{eqnarray}
b_0(t) & \approx & \left(\frac{\alpha+2}{2}t\right)^{2/(\alpha+2)}\nonumber\\ 
& &\times \left(\frac{2 \ln t}{\alpha+2} \left[ \frac{\alpha + 2}
{16 F_m} + \frac{3}{32 F_m}\right]\right)^{\alpha/(\alpha + 2)}, 
\label{BTEQ}
\end{eqnarray}   
whence,
\begin{eqnarray}
\frac{p_0^2(t)}{b_0(t)} & \approx & \frac{\ln t}{2 F_m} \left[1 + 
\frac{3}{2 (\alpha + 2)} \right]\,,\nonumber \\
& & \nonumber \\ 
a(t)& \approx & \left(\frac{4}{(\alpha+2)^2} \frac{\ln t}{t}\right)
^{1/(\alpha + 2)}\nonumber \\
& &\times \left[\frac{\alpha + 2}{16 F_m} + 
\frac{3}{32 F_m}\right]^{1/(\alpha + 2)}\,.
\label{APTEQ}
\end{eqnarray} 
The above results for $a(t)$, $b_0(t)$ and $p_0(t)$ justify our original 
ansatz. From (\ref{chals}) we can define the characteristic length 
scales in three directions:
$L_x = \gamma(t^2 b_0/p_0)^{1/2} \sim \Upsilon(\alpha) 
\,\gamma\,(t^{2\alpha +5}/\ln t)^{1/2(\alpha + 2)}$, 
and $L_y = L_z = (b_0/p_0)^{1/2} \sim \Upsilon(\alpha) 
\,(t/\ln t)^{1/2(\alpha + 2)}$, by setting $u=k_xL_x$, $v=k_yL_y$, and 
$w=k_zL_z$, where 
\begin{equation}
\Upsilon^{-1}(\alpha) = \sqrt{2} \left(\frac{4}{(\alpha + 2)^2} 
\left[\frac{\alpha + 2}{16 F_m} + \frac{3}{32 F_m}\right]
\right)^{1/2(\alpha + 2)}\,.
\label{COEP}
\end{equation} 
Exact values for $F_m(\alpha)$ and 
$(\,u_m(\alpha),v_m(\alpha),w_m(\alpha)\,)$ can be found by specifying 
values of $\alpha$. For example, values for $\alpha = 1$ and 2 are shown in 
table 1, while  for $\alpha = 0$, results presented in\cite{Rapapa} are 
recovered.

Using (\ref{FUS}), (\ref{SECOS}), and (\ref{APTEQ}), it is easy to show that 
the structure factor becomes
\begin{equation}
S({\bf k},t) = {\rm Const.}\,\left[\ln V_s\right]^{3/2} 
V_s^{F({\bf q})/F_m}\,,
\label{EXACS}
\end{equation}
with scaled momentum  ${\bf q}$ and 'scale volume' $V_s$ given by 
\begin{eqnarray}
{\bf q} &=& (k_xL_x,k_yL_y,k_zL_z)\,,\nonumber \\
 V_s  &=& L_xL_yL_z \sim \gamma t^{(7+2\alpha)/2(\alpha + 2)}
/(\ln t)^{3/2(\alpha + 2)}\,,
\end{eqnarray}
respectively. Eq.(\ref{EXACS}) exhibits multiscaling behavior 
(i.e the power of the 'scale volume' depends continuously on the scaling 
variables). We anticipate, however, that for finite $n$, standard 
scaling will be recovered and the $\ln t$ terms (which appear in the 
characteristic length scales) will be absent as has been shown 
explicitly for the case with both $\alpha$ and $\gamma$ = 0\cite{Humayun}.

\section{Results} 
Both the positions of the stationary points of $F(u,v,w)$ and its value 
depend on $\alpha$ ( apart from points (0,0,0) and 
$\pm(0,0,1/\sqrt{2})$ ), see table 1. The global picture of $S({\bf k},t)$ 
is determined by $F(u,v,w)$ (i.e. the stationary points and their type 
together with values of  $F(u,v,w)$ at stationary points) because 
$\ln S({\bf k},t) = (F(u,v,w)/F_m) \ln V_s$ 
[plus $\bf k$-independent terms]. The features of $F(u,v,w)$ 
for $\alpha = 1$ and 2 are summarised in table 1 and table 2. 

The parallel ridges found in Figure 1 are similar to ones found in 
experiments\cite{Beysens}. The global maxima are 
connected by almost straight ridges to the 'inflexion maxima'. As the 
value of $\alpha$ increases, the height of each ridge decreases towards the 
limiting value $F = 1/4$ (and also the length of each ridge increases) 
while the line connecting the two saddle points and the minimum approaches 
the line $u = -v$ with $F = 0$. In the $(k_x,k_y)$ plane, the ridges 
become narrower, higher, and closer together as a function of $\bf k$ with 
increasing time $t$. The angle $\theta$ between the ridges and the shear 
direction ($k_y$ direction in this case) is a good measure of theory against 
experiments because the depth of the temperature quench is unimportant as 
far as the time dependence is concerned. We find
\begin{equation}
\tan (\theta) = \frac{C_2(\alpha)}{\gamma t}\,,
\label{theta}
\end{equation}  
where $C_2(\alpha)$ is a constant which depends on $\alpha$, e.g.  
$C_2(0) = 2(1-1/\sqrt{6}), C_2(2) = (7-\sqrt{7})/4$.
Eq. (\ref{theta}) implies that in the $(k_x,k_y)$ plane, the ridges move 
closer to the shear direction as time increases, this behavior is found 
both in simulations\cite{Gonnella} and experiments\cite{Lauger}.

\begin{figure}
\narrowtext
\vspace{-3.0cm}
\centerline{\epsfxsize\columnwidth\epsfbox{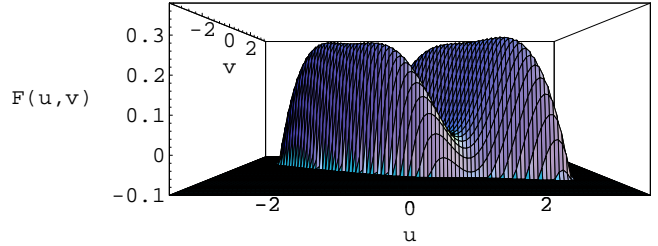}}
\vspace{-2cm}
\caption{The graph of $F(u,v,0)$ for $\alpha = 1$. Values for $F(u,v,0) 
< -0.1$ are not shown.} 
\label{fuv}
\end{figure}

\begin{figure}
\narrowtext
\vspace{-1.85cm}
\centerline{\epsfxsize\columnwidth\epsfbox{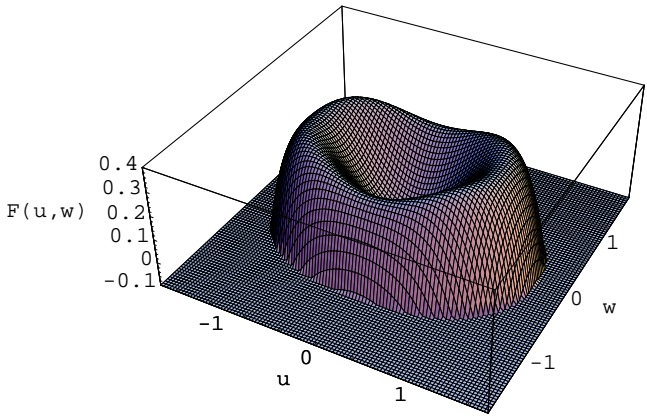}}
\vspace{-0.8cm}
\caption{The graph of $F(u,0,w)$ for $\alpha = 1$. Values for $F(u,0,w) 
< -0.1$ are not shown.} 
\label{fuw}
\end{figure}

In the $(u,w)$ plane, there are four maxima, four saddle points and the 
global minimum at the origin. These are shown in table 2 (for $\alpha = 1, 
2$), and they can easily be seen in Figure \ref{fuw} for $\alpha = 1$. When 
$\alpha$ increases, the peaks and the higher saddle point (i.e saddle point 
with higher value of $F(u,0,w)$ reduces towards the limiting value 
$F = 1/4$ (i.e $F_m (u,0,w)$ is deformed towards a ring of radius 
$1/\sqrt{2}$). For each value of $\alpha$, the structure factor pattern 
in the $(k_x,k_z)$ plane will decrease faster in the flow (i.e $k_x$) 
direction as a function of $t$, resulting in an elliptical shape with 
major axis along $k_z$ direction. This elliptical shape has been observed 
in experiments\cite{Lauger} for scalar fields while the four peaks were not 
observed.

The excess viscosity $\Delta\eta$, and normal stresses $\Delta N_1$, 
$\Delta N_2$ derived by Onuki\cite{Onuki} can be evaluated in the 
asymptotic limit:
\begin{eqnarray}
\Delta\eta &=& -(1/\gamma)\int\,\frac{d^3k}{2\pi)^3}\,k_xk_y\,S({\bf k},t)
\,,\nonumber \\
  &\sim &\frac{1}{\gamma^2}\,\left(\frac{\ln t}{t^{\alpha +3}}
\right)^{1/(\alpha + 2)}\,,\nonumber \\
\Delta N_1 &=& \int\,\frac{d^3k}{2\pi)^3}\,(k_y^2 - k_x^2)\,S({\bf k},t)
\,,\nonumber \\
  &\sim &\left(\frac{\ln t}{t}\right)^{1/(\alpha + 2)}
\,,\nonumber \\
\Delta N_2 &=&\int\,\frac{d^3k}{2\pi)^3}\,(k_y^2 - k_z^2)\,S({\bf k},t)
\,,\nonumber \\
  &\sim &\left(\frac{\ln t}{t}\right)^{1/(\alpha + 2)}
\,.\nonumber \\
\label{VNN}
\end{eqnarray}
Numerical calculations\cite{Suppa} show that both excess viscosity 
$\Delta\eta(t)$, and the normal stress $\Delta N_1(t)$ reach a peak 
before the asymptotic scaling result. It is not possible to realise 
this effect as our calculations are strictly valid in the 
asymptotic regime.

\medskip

\begin{center}
\begin{tabular}{|c|c|c|c|c|}
\hline
\multicolumn{5}{|c|}{$\alpha = 1$}\\
\hline
Position (u,w) & No.  & F & Value & Type \\ 
\hline
$(0,0)$ & 1 &   0         & 0 & Min \\ \hline
$\pm(\sqrt{55/6}/3,0)$ & 2 & 55/168 & .32738 & S \\ \hline
$\pm(0,1/\sqrt{2})$ & 2 & 1/4 & .25 & S \\ \hline
$\pm(\sqrt{110/129},\pm\sqrt{5/43})$ & 4 & 100/301 & 0.3322&Max \\ 
\hline
\multicolumn{5}{|c|}{$\alpha = 2$}\\
\hline
Position (u,w) & No.  & F & Value & Type \\ 
\hline
$(0,0)$ & 1 &   0  & 0 & Min \\ \hline
$\pm(\sqrt{7/2}/2,0)$ & 2 & 14/45 & .31111& S \\ \hline
$\pm(0,1/\sqrt{2})$ & 2& 1/4 & .25 & S \\ \hline
$\pm(\sqrt{35/3}/4,\pm 1/3)$&4&17/54&0.31481&Max \\ 
\hline
\end{tabular}
\end{center}

\underline{Table 2.} Stationary points of $F(u,0,w)$: Max = maximum, 
Min = minimum, S = saddle point. See also FIG. 2.

\medskip

In analogy to 2-$D$ numerical calculations presented in\cite{Suppa}, it 
is easy to repeat the calculation for  2-$D$ in the shear-flow 
($k_x$ , $k_y$) plane. The self-consistent equation for $a(t)$ leads to 
(with $C_3(\alpha)$ constant) 
\begin{equation}
1 = \frac{C_3(\alpha)\Delta}{\gamma t p_0}\exp
\left(2\,\frac{p_0^2}{b_0}\,F^{2D}_m(\alpha)\right)\,,
\label{SECTD}
\end{equation}
from which to leading order in $t$,
\begin{eqnarray} 
b_0(t) & \approx & \left(\frac{\alpha+2}{2}t\right)^{2/(2+\alpha)}\nonumber\\ 
& &\times \left(\frac{2 \ln t}{\alpha+2} \left[ \frac{\alpha + 2}
{16 F^{2D}_m} + \frac{1}{16 F^{2D}_m}\right]\right)^{\alpha/(\alpha + 2)}
\,,\nonumber\\
& &  \nonumber\\
\frac{p_0^2(t)}{b_0(t)} &\approx&  \frac{\ln t}{2 F^{2D}_m} \left[1 + 
\frac{1}{(\alpha + 2)} \right]\,,\nonumber \\
 & &  \nonumber\\
a(t)& \approx & \left(\frac{4}{(\alpha+2)^2} \frac{\ln t}{t}\right)
^{1/(2+\alpha)}\nonumber \\ 
&  &\times\left[\frac{\alpha + 2}{16 F^{2D}_m} + 
\frac{1}{16 F^{2D}_m}\right]^{1/(\alpha + 2)}\,.
\label{APTEQTD}
\end{eqnarray} 

Using both (\ref{SECTD}) and (\ref{APTEQTD}), the structure factor can be 
written down as $S({\bf k},t) = {\rm Const}. \,(\ln A_s) 
A_s^{F({\bf q})/F^{2D}_m}$, where $A_s = L_x L_y$ is the 'scale area' and 
${\bf q} = (k_xL_x , k_yL_y)$ with $L_x \sim \gamma (t^{5+2\alpha}/\ln t)
^{1/2(\alpha + 2)}$, $L_y \sim (t/\ln t)^{1/2(\alpha + 2)}$ 
and ${F}^{2D} = F(u,v,0)$. Therefore the structure factor pattern is similar 
to Figure 1. The oscillations between the peaks\cite{Suppa} which terminate 
the two parallel ridges, we believe are suppressed in the long time regime 
(i.e. they are the preasymptotic decaying transients).

\section{Summary and Remarks}
We have analytically studied the effect of both shear and 
order-parameter-dependent mobility on phase-separation within the 
large-$n$ limit. Shear introduces anisotropy in the structure factor 
pattern because of different growth rates in the flow direction and 
directions perpendicular to flow. At fixed time $t$ and ${\bf k}$, 
$\alpha$ distorts the shape of the structure factor $S({\bf k},t)$ 
(this is evident in the $F(u,v,0)$ and $F(u,0,w)$ patterns). Similar 
to all studies previously done, the order-parameter-dependent mobility 
slows down the rate of coarsening 
(i.e. $L_i(\alpha =  0) > L_i(\alpha \ne 0)$, where $i = x$, $y$, $z$). 
We believe the multiscaling found here to be the result 
of the large-$n$ approximation, and that for any finite $n$, standard 
scaling will be obtained with the same characteristic length scales but 
without the $\ln t$ terms,  
$L_x \sim t^{(2\alpha+5)/2(\alpha+2)}$, 
$L_y \sim t^{1/2(\alpha+2)}$, and 
$L_z \sim t^{1/2(\alpha+2)}$. 
The excess viscosity $\Delta\eta(t)$, and the normal stresses 
(i.e $\Delta N_1(t)$ and $\Delta N_2(t)$) relax to zero as 
$(\ln t/t^{\alpha + 3})^{1/(\alpha + 2)}$ and $(\ln t/t^{1/(\alpha + 2)}$ 
respectively in the scaling limit. Again we expect logarithmic terms to be 
absent for finite $n$.

\bigskip

\begin{center}
\begin{small}
{\bf ACKNOWLEDGEMENTS}
\end{small}
\end{center}
NR thanks A. Bray for suggesting this problem, and for discussions.
This work was supported by the Commonwealth Scholarship Commission.

\end{multicols}

\begin{references}  
\bibitem{Bray} A. J. Bray, Adv.\ Phys.\ {\bf 43}, 357 (1994). 
\bibitem{Onuki} A. Onuki, J. Phys.: Condens. Matter 
{\bf 9}, 6119 (1997), and references therein.
\bibitem{Rothman}D. H. Rothman, Phys.\ Rev.\ Lett.\ {\bf 65}, 3305 (1990). 
\bibitem{Padilla}P. Padilla and S. Toxvaerd, J. Chem.\ Phys.\ 
{\bf 106}, 2342 (1997).
\bibitem{Yeomans}A. J. Wagner and J. M. Yeomans, preprint 
(cond-mat/9904033).
\bibitem{Gonnella}F.Corberi, G. Gonnella, and A. Lamura, preprint 
(cond-mat/9904423).
\bibitem{Rapapa}N. P. Rapapa and A. J. Bray, preprint (cond-mat/9904396).
\bibitem{Corberi} F. Corberi, G. Gonnella, and A. Lamura, 
Phys.\ Rev.\ Lett.\ {\bf 81}, 3852 (1998).
\bibitem{Beysens}C. K. Chan, F. Perrot, and D. Beysens, 
Phys.\ Rev.\ A {\bf 43}, 1826 (1991). 
\bibitem{Hashimoto}T. Hashimoto, K. Matsuzaka, E. Moses, and A. Onuki, 
Phys.\ Rev.\ Lett.\ {\bf 74}, 126 (1995). 
\bibitem{Lauger}J. L\"auger, C. Laubner, and W. Gronski,
Phys.\ Rev.\ Lett.\ {\bf 75}, 3576 (1995).
\bibitem{Langer}J. S. Langer, M. Bar-on, and H. D. Miller, Phys.\ Rev. A,\ 
{\bf 11}, 1417 (1975); K. Kitahara and M. Imada, Prog.\ Theor.\ Phys.\ 
Suppl.\ {\bf 64}, 65 (1978).
\bibitem{Jasnow}D. Jasnow, in {\em Far from Equilibrium Phase 
Transitions} ed. L. Garrido ( Springer-Verlag, Berlin, 1988); K. Kitahara, 
Y. Oono, and D. Jasnow, Mod.\ Phys.\ Lett.\ B\ {\bf 2}, 765 (1988).
\bibitem{Puri}S. Puri, A. J. Bray, and J. L. Lebowitz, Phys.\ 
Rev.\ E\ {\bf 56}, 758 (1997); A. M. Lacasta, A. Hernandez-Machado, 
and J. M. Sacho, Phys.\ Rev.\ B\ {\bf 45}, 5276 (1992).
\bibitem{Castellano}F. Corberi and C. Castellano, Phys.\ Rev.\ E\ 
{\bf 58},4658 (1998). 
\bibitem{Emmbra}A. J. Bray and C. L. Emmott, Phys.\ Rev.\ B\ {\bf 52}, 
R685 (1995).
\bibitem{Emmott}C. L. Emmott and A. J. Bray, Phys.\ Rev.\ E\ {\bf 59}, 
213 (1999).
\bibitem{Ahluwalia}R. Ahluwalia, Phys.\ Rev.\ E\ {\bf 59}, 263 (1999).
\bibitem{Coniglio}A. Coniglio and M. Zannetti, Europhys.\ Lett.\ 
{\bf 10}, 575 (1989).
\bibitem{Humayun}A. J. Bray and K. Humayun, Phys.\ Rev.\ Lett.\ 
{\bf 68}, 1559 (1992).
\bibitem{Ruggiero}A. Coniglio, P. Ruggiero, and M. Zannetti 
Phys.\ Rev.\ E\ {\bf 50}, 1046 (1994); F. Corberi, A. Coniglio, and 
M. Zannetti, Phys.\ Rev.\ E {\bf 51}, 5469 (1995); C. Castellano and 
M. Zannetti, Phys.\ Rev.\ Lett.\ {\bf 77}, 2742 (1996); 
Phys.\ Rev.\ E {\bf 53}, 1430 (1996); Phys.\ Rev.\ E {\bf 58}, 5410 (1998).
\bibitem{Suppa}G.Gonnella, A. Lamura, and D. Suppa, preprint 
(cond-mat/9904422). 
\end{references}
\end{document}